\documentclass[amssymb,useAMS,prd,aps,amsmath,nofootinbib,superscriptaddress,twocolumn]{revtex4}

\newcommand{\kumiko}{\color{black}}
\newcommand{\joe}{\color{black}}
\newcommand{\celine}{\color{black}}
\usepackage{graphicx}
\usepackage{amssymb}
\usepackage[caption=false]{subfig}
\usepackage{array, xcolor}
\usepackage[colorlinks=true,linkcolor=blue,citecolor=blue]{hyperref}

\begin{document}
\title{Shocking Signals of Dark Matter Annihilation}
\author{Jonathan H. Davis}
\affiliation{Institut d'Astrophysique de Paris, UMR 7095, CNRS, UPMC Universit\'{e} Paris 6, Sorbonne Universit\'{e}s, 98 bis boulevard Arago, 75014 Paris, France}
\author{Joseph Silk}
\affiliation{Institut d'Astrophysique de Paris, UMR 7095, CNRS, UPMC Universit\'{e} Paris 6, Sorbonne Universit\'{e}s, 98 bis boulevard Arago, 75014 Paris, France}
\affiliation{Department of  Physics \& Astronomy, The Johns Hopkins University \\
3400 N Charles Street, Baltimore, MD 21218, USA}\affiliation{Laboratoire AIM-Paris-Saclay, CEA/DSM/IRFU, CNRS, Universite Paris Diderot,  F-91191 Gif-sur-Yvette, France}
\affiliation{Beecroft Institute of Particle Astrophysics and Cosmology, Department of Physics,
University of Oxford, Denys Wilkinson Building, 1 Keble Road, Oxford OX1 3RH, UK}
\author{C\'eline B\oe hm}
\affiliation{Institute for Particle Physics Phenomenology, Durham University, Durham, DH1 3LE, UK}
\affiliation{LAPTH, U. de Savoie, CNRS,  BP 110,
  74941 Annecy-Le-Vieux, France}
\author{Kumiko Kotera}
\affiliation{Institut d'Astrophysique de Paris, UMR 7095, CNRS, UPMC Universit\'{e} Paris 6, Sorbonne Universit\'{e}s, 98 bis boulevard Arago, 75014 Paris, France}
\author{Colin Norman}
\affiliation{Department of  Physics \& Astronomy, The Johns Hopkins University \\
3400 N Charles Street, Baltimore, MD 21218, USA}
\affiliation{Space Telescope Science Institute, 3700 San Martin Drive, Baltimore, MD 21218, USA}
\begin{abstract}
{We examine whether charged particles injected  by self-annihilating Dark Matter into regions undergoing Diffuse Shock Acceleration (DSA) can be accelerated to high energies.
We consider three astrophysical sites where shock acceleration is supposed to occur, namely the Galactic Centre, galaxy clusters and Active Galactic Nuclei (AGN).  For the Milky Way,   we find that the acceleration of cosmic rays injected by dark matter could lead to a bump in the cosmic ray spectrum provided that the product of the efficiency of the acceleration mechanism 
and  the concentration of DM particles is high enough. Among  the various acceleration sources that we consider (namely supernova remnants (SNRs),  Fermi bubbles and AGN jets), we find that the Fermi bubbles are a potentially  more efficient accelerator than SNRs. However both could in principle accelerate electrons and protons injected by dark matter to very high energies. At the extragalactic level, the acceleration of dark matter annihilation products could  be responsible for enhanced radio emission from colliding clusters   and prediction of an increase of the anti-deuteron flux generated near AGNs. }
\end{abstract}
\maketitle

\section{Introduction}
Cosmic rays {\kumiko are detected up to $\gtrsim 10^{20}$~eV} energies and are composed of a variety of particles, which, depending on the energy, could be electrons, positrons, hadrons and nuclei~{\kumiko \cite{PDG,B90,Gaisser91,Stanev10,KO11,Letessier11}.}
The existence of these high-energy particles requires an astrophysical acceleration mechanism to ultra-relativistic energies.
One popular process is first-order Fermi acceleration, in which charged particles are injected in magnetized shock regions. Thanks to scattering on magnetic field inhomogeneities, particles can repeatedly cross the shocks and gain energy. This process, also known as Diffuse Shock Acceleration (DSA)~\cite{Kang:2012nz,2014MNRAS.437.2291V,Bykov:2012ca,Rieger:2006md}, is meant to accelerate particles, such as electrons and protons, to high energies at a variety of astrophysical galactic and extragalactic sites.

{While DSA  has been advocated as a possible acceleration mechanism  in our own Galaxy, the main astrophysical site where the acceleration of cosmic rays can take place is not yet established. The most plausible source appears to be supernovae remnants (SNRs) and is consistent with X- and gamma-ray data for electron and hadron acceleration respectively \citep{Tavani10,Giuliani11, 2013Sci...339..807A}. However the Fermi bubbles could be another powerful accelerator in our galaxy. }

{ Assuming  no other source of  cosmic rays  other than the thermal population, the amount of energy required to explain the observed  spectrum could be a possible issue for SNRs: protons would take up to 30\% of the SNR shock energy while electrons would only take 1\%  \cite{2001RPPh...64..429M}.  Another possible issue lies in the excess 
of leptonic cosmic rays at high energy. Several experiments such as PAMELA, Fermi-LAT, HESS, MAGIC, and AMS-02 \cite{Adriani09,Adriani11,Adriani2013,Abdo09_elec,Ackermann10,Ackermann12,Aharonian09_HESS,Blum2013,Tridon11} have indeed collected leptonic cosmic rays up to around TeV energies. However these observations are difficult to explain with thermally-injected cosmic rays and the standard DSA mechanism due to energy losses of cosmic rays in the Galaxy. Serious consideration of the details of the acceleration mechanism  of cosmic rays in the Milky Way may therefore be useful.}

{Similarly, the radio emission from the so-called `Toothbrush' relic in the cluster 1RXS J0603.3+4214 ~\cite{2012MNRAS.425L..76B,Vazza:2013hma} and the non-thermal spectra from several AGNs indicate that shock-acceleration is also occurring in extragalactic objects and the observed spectrum might also require introduction of a new population of cosmic rays.}

{Here, we entertain the idea that dark matter (DM) self-annihilations sustain a non-thermal source of cosmic rays that get accelerated to very high energies (well above the DM mass threshold) thanks to  astrophysical shocks. This hypothesis is justified by the fact that both galactic and extragalactic sites possess high number densities of DM in their centres, where astrophysical accelerators also are located~\cite{Newman:2012nw,Gondolo:1999ef}, and could therefore explain cosmic ray observations. }
\textcolor{black}{This also solves a long-standing problem with injection into shocks, namely how the injected particles build in energy from non-relativistic velocities. In our case the particles are already injected at relativistic energies by  dark matter annihilations.}

The combination of both DM injection and DSA { gives rise to an  interesting signature}. 
Dark matter injection alone gives an energy spectrum (of electrons, protons or other particles) bounded from above by the DM mass.
{In} contrast, shock acceleration can bring injected particles to much higher energies, producing a power-law distribution over all energies.
Indeed in section~\ref{sec:model}, we show that when { a DM  contribution is added to the ordinary cosmic-ray component}, the power-law behaviour common to DSA is maintained, but with a low-energy cut-off set by the DM mass. This cut-off  is less prominent if the energy loss rate is large. In section~\ref{sec:super} we discuss the potential for DM injection towards the Galactic Centre to create observable signatures in cosmic ray data and radio measurements, and in section~\ref{sec:extra} we discuss extra-galactic signals. We conclude in section~\ref{sec:conc}.

\section{Empirical Model for Dark Matter Injection at Shocks \label{sec:model}}
{In this section, we present the  model that we implement to describe the shock acceleration of particles injected by dark matter. Our mechanism is based on Fermi acceleration.  First we assume that cosmic rays are steadily injected  by DM annihilations (see Eq.~\ref{eq:DMinj}). Next we assume  that a fraction of these particles, $\epsilon$,  ends up in the shock region, where they  get trapped by the magnetic field $B$. The time-scale over which they are expected to stay in the acceleration region  is defined by $T_{\mathrm{trap}} \sim E / Z e B c v_s$, where $Z e$ is the electric charge of the trapped particle~\cite{Achterberg:2001rx} and $v_s$ is the speed of the shock. During this time the particles may gain energy by crossing the shock and also lose energy through standard processes, namely inverse Compton and synchrotron losses for the leptonic components  and  pion or $e^+ e^-$ pair production for the hadronic part.  }

Since the shock occurs within a much smaller volume than that characteristic of the  injection of dark matter, the efficiency factor $\epsilon$ can be as small as $\sim 10^{-5}$~\cite{2014MNRAS.437.2291V}. This small value could be however compensated by a large DM number density near the acceleration site. Besides, even such a very small value can give rise to the observed electron-proton ratio  via DSA \cite{2014ApJ...788..142K}. 
\begin{figure}[t]
\centering
\subfloat{ \includegraphics[width=0.48\textwidth]{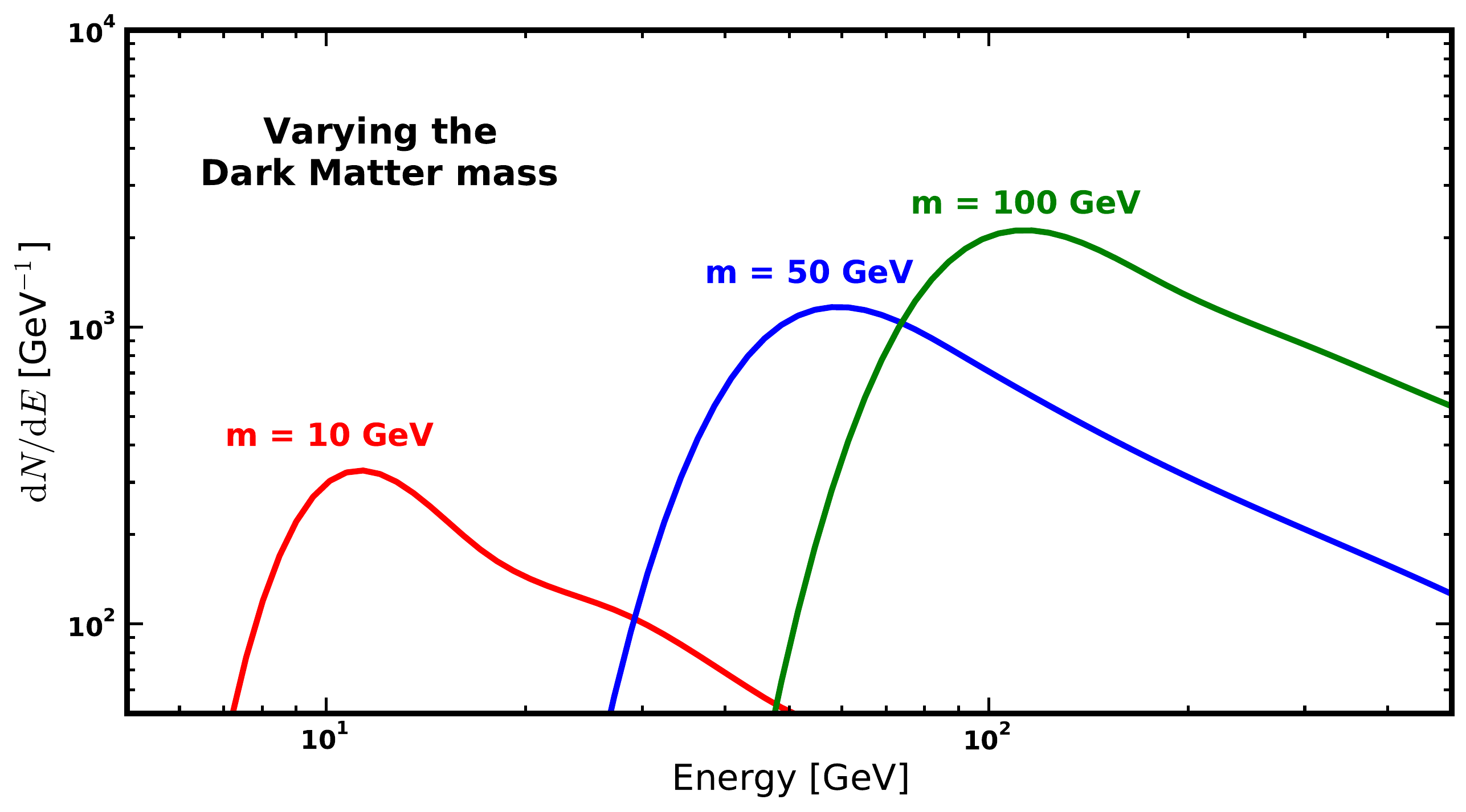}} \\
\subfloat{ \includegraphics[width=0.48\textwidth]{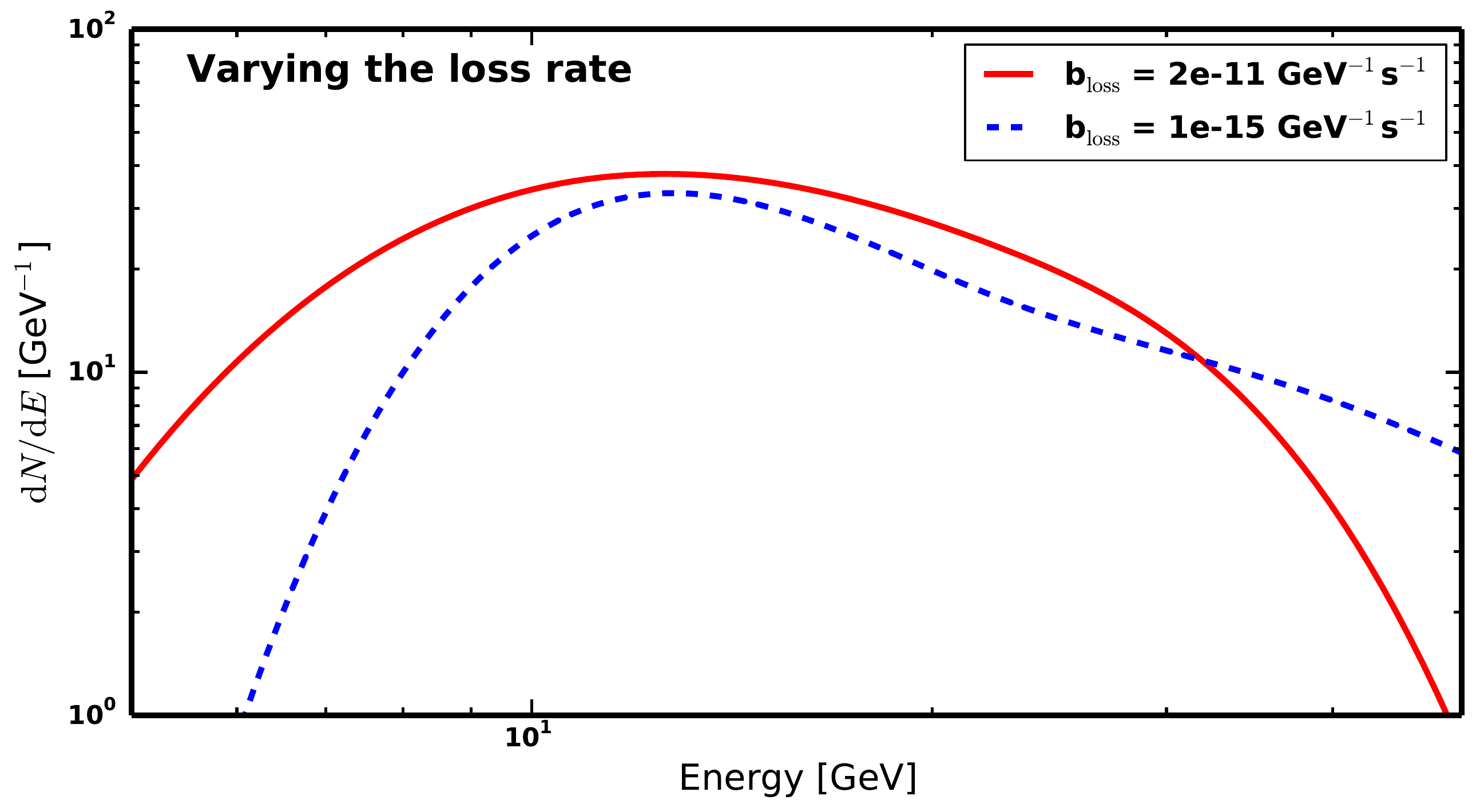}}
\caption{Spectra of electrons which have been injected into a region of Diffuse Shock Acceleration by self-annihilating Dark Matter. \textbf{Upper panel:} Dependence of the electron spectrum within the shock region on the Dark Matter mass. \textbf{Lower panel:} Variation of the electron spectrum with the energy loss rate, assuming a DM mass of 10~GeV.}
\label{fig:shock_nonrel}
\end{figure}

{ In this work, we will assume a  delta-function for  the spectrum of electrons or protons injected by the DM into the shock region. This is equivalent to assuming that the spatial diffusion can be neglected and that the trapping of cosmic rays is much faster than the energy losses.}

{\kumiko The physics inside the shock 
can be modeled roughly using the Fermi mechanism.} We give the details of our empirical model in appendix~\ref{app1}.
The evolution of the particle spectrum (per unit volume), $\mathrm{d}n / \mathrm{d}E$, over time $t$ is modeled by taking discrete time-steps $\Delta t$.
The spectrum of particles still in the shock region by the end of the time-step $\Delta t$ is given by

\begin{eqnarray}
\frac{\mathrm{d}n}{\mathrm{d}E}_{\mathrm{final}} &=& \mathrm{exp} \left[ - \frac{\Delta t}{T_{\mathrm{trap}}} \right] \frac{\mathrm{d}n}{\mathrm{d}E}(E- b_{\mathrm{loss}} E^2(t) \Delta t) \nonumber \\
&+& \beta \cdot \left(1 - \mathrm{exp} \left[ - \frac{\Delta t}{T_{\mathrm{trap}}} \right] \right) \nonumber \\ &\cdot&  \frac{\mathrm{d}n}{\mathrm{d}E}(E- b_{\mathrm{loss}} E^2(t) \Delta t + \delta \cdot E) ,
\end{eqnarray}
where the first term represents the particles which remain trapped without crossing the shock and the second gives the spectrum of the accelerated particles. {\kumiko The term $b_{\mathrm{loss}}$ is the total loss rate, and the factors $\beta$ and $\delta$ the probability for a particle to return to the shock, and the fractional energy gain per shock crossing, respectively. These three parameters depend on the nature of the particles and the properties of the acceleration region (see App.~\ref{app1}).}

{After many time-steps, the number of particles inside the shock region reaches a steady-state: the number of particles that escape and those injected compensate.} For fixed $\beta$ and $\delta$ the form of the resulting steady-state spectrum depends on the injection spectrum  $f_{\mathrm{DM}} (E,m)$ and the energy loss rate $b_{\mathrm{loss}}$. {Since we took a delta function for $f_{\mathrm{DM}} (E,m)$, we expect all the annihilation products injected by DM to have their energy shifted above the DM mass threshold. }

 {\celine In figure~\ref{fig:shock_nonrel}  we show the electron (and positron) spectrum injected by DM annihilation after DSA. While the initial $e^+ e^-$  spectrum at injection has a cut-off above the DM mass, the effect of DSA is to significantly alter the spectrum by accelerating the  cosmic rays above the DM mass threshold. As a result, the final (accelerated) spectrum contains a  low-energy cut-off fixed by the DM mass and, above this cut-off, the spectrum follows a power-law (see the upper panel of figure~\ref{fig:shock_nonrel}), consistent with mono-energetic injection scenarios such as considered in ref.~\cite{1984A&A...136..227S}. }

 Our results have been obtained using a simple empirical simulation. We have assumed that the physics of the acceleration mechanism does not depend  on the exact location of the particle injection and have also neglected the spatial dependence of the energy-loss rate. Even so, the broad features that we investigate in detail in the following, such as the low-energy cut-off around the DM mass, should remain in a more robust simulation.

\section{Accelerated cosmic rays in the Galactic Centre \label{sec:super}}

{We now discuss the case of cosmic rays injected by DM annihilations in the Milky Way centre. We will consider two acceleration sources, namely SNRs and  the Fermi bubbles. }

\subsection{Acceleration mechanism}

{We will focus on the galactic centre where the DM energy density is expected to be the largest.  The number  (and properties)  of SNRs near the centre are unknown so the results highlighted in this section are only for illustrative purposes. However these results allow us to determine which of the Fermi bubbles and SNRs are the most powerful accelerator of  electrons and protons.}

For electrons, we assume that the major energy loss mechanism is synchrotron radiation due to the strong magnetic field (we assume that the loss rate from inverse Compton scattering is $b_{\mathrm{loss}}^{\mathrm{IC}} \approx 2.5 \cdot 10^{-17} \mathrm{GeV}^{-1} \mathrm{s}^{-1}$ due to scattering off the Cosmic Microwave Background {and interstellar field components}~\cite{2013ApJ...768..106S} {and thus neglect it with respect to synchrotron losses\footnote{Such an assumption is only justified if the magnetic field is strong enough but large values are justified for the sites we consider.}}).   As an illustration, we take the magnetic field near the galactic centre to be $B = 50$~$\mu$G based on the upper limit derived in~\cite{Crocker:2010qn}. Hence assuming that the shock speed is {\celine $v_s \approx 3 \cdot 10^6$~m \ s$^{-1}$~\cite{Hoeft:2006fg} } we find
\begin{eqnarray}
T_{\mathrm{trap}} &\sim& 10^{-2} \left[\frac{E}{\mathrm{GeV}} \right]  \left[\frac{B}{\mathrm{G}} \right]^{-1} \mathrm{s} = 200 \cdot   \left[\frac{E}{\mathrm{GeV}} \right] \, \mathrm{s} \\
b_{\mathrm{loss}}^{\mathrm{sync}} &\approx& 2.54 \cdot 10^{-18} \left[\frac{B}{\mathrm{\mu G}} \right]^{2}  \mathrm{GeV}^{-1} \mathrm{s}^{-1}.
\end{eqnarray}
\textcolor{black}{Hence for an electron with energy of 1~GeV the average life-time is $\sim 6 \cdot 10^6$~years.}
For protons we assume that the energy losses are dominated by the production of $e^+ e^-$ pairs {(from proton interactions with photons)}  for which the energy loss rate is roughly three orders of magnitude smaller. Additionally we need to make an assumption on the distribution of supernovae around the galactic centre and on the spread of the Mach number $\mathcal{M}$ of supernova shocks. For our estimate we assume that all supernovae shocks have a Mach number of $\mathcal{M} \approx 10$~\cite{2011Ap&SS.336..257R}. {The assumptions made here are likely to be too simple and could lead us to overestimate  the flux. However we are only interested in showing how this mechanism could explain the observed high energy cosmic ray spectrum in the galaxy. }

\subsection{Acceleration by SNRs} 
{To estimate the number of DM-injected products which could be accelerated by shocks in SNRs, we assume that the average SNR shock sweeps out a sphere of radius 30~pc and lasts for $\sim 10$~years with a rate of supernova  explosions $\sim 0.01$ per century~\cite{Crocker:2010qn}. This gives an effective total volume for the supernovae  of $V_{\mathrm{SN}} \sim (4 \pi /3) \cdot (30 \mathrm{pc})^3 \cdot 0.1 \, \mathrm{centuries} \, \cdot  0.01 \, \mathrm{SN}/\mathrm{century} \sim 100 \, \mathrm{pc}^3 $. Such a volume has to be compared with that for DM annihilations in the Galactic Centre (GC), $V_{\mathrm{GC}}$,  which we define as being contained in  a sphere of radius 1 kpc centred on the GC. }

 {The  fraction of cosmic rays injected in the acceleration region} is estimated to be,
\begin{equation}
R_{\mathrm{DM}}^{\mathrm{SN}} \sim \epsilon \mathcal{B} \left[ \frac{V_{\mathrm{SN}}}{V_{\mathrm{GC}}} \right] \, \int_{{V_{\mathrm{GC}}}} \mathrm{d}^3 r \, n_{\mathrm{DM}}^2(r) \langle \sigma v \rangle,
\end{equation}
where $\epsilon$ is the efficiency factor accounting for the proportion of cosmic rays trapped in the acceleration region ~\cite{Vazza:2013hma} and  $\mathcal{B}$ is
the boost factor accounting for an increase in the DM energy density ($\rho_{\mathrm{DM}}$) with respect to a Navarro-Frenk-White (NFW) profile~\cite{Navarro:1996gj}. 
The number density $n_{\mathrm{DM}}$ is related to the energy density by $\rho_{\mathrm{DM}} = n_{\mathrm{DM}} \ m_{\mathrm{DM}}$.

\begin{figure}[t]
\centering
\subfloat{ \includegraphics[width=0.48\textwidth]{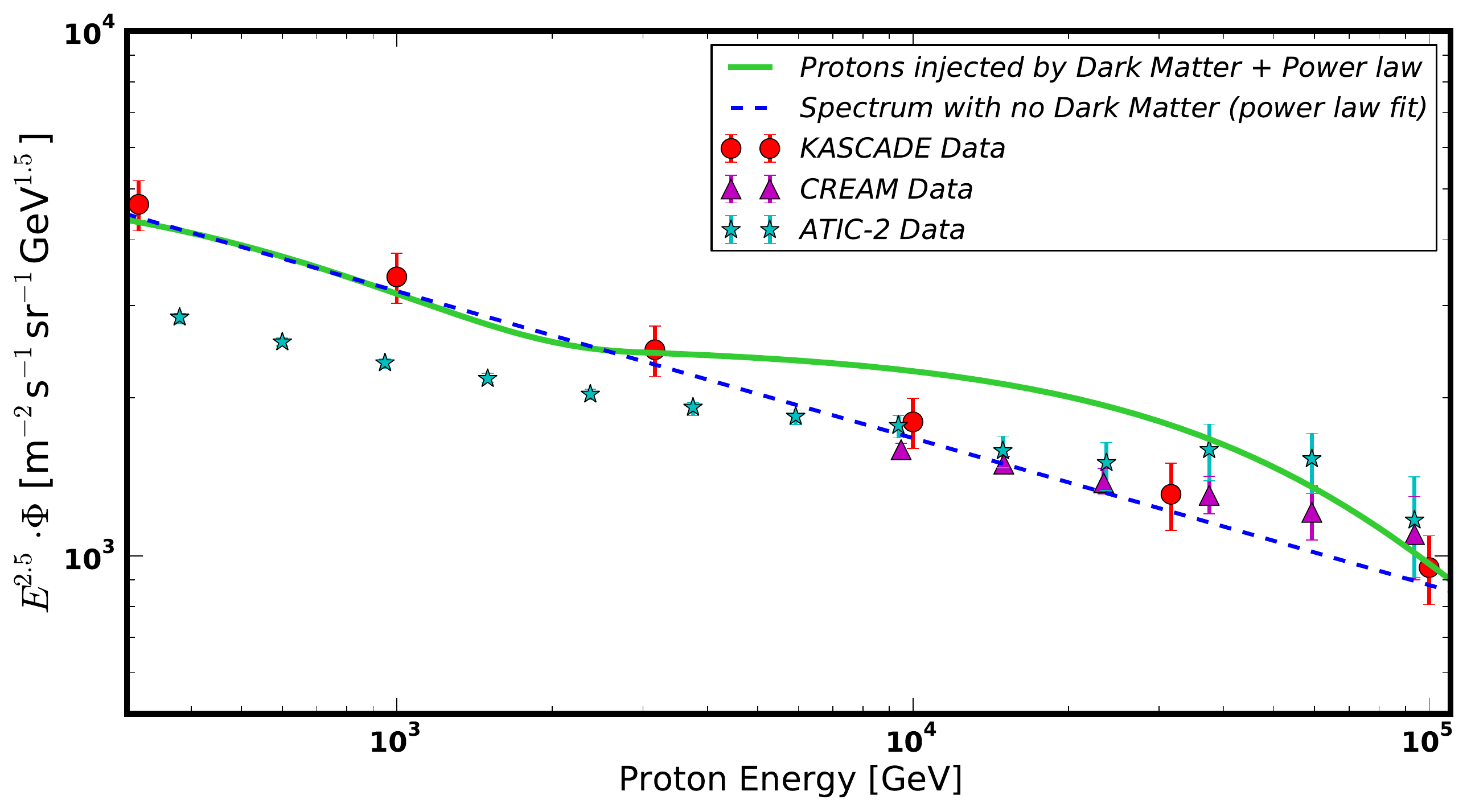}}
\caption{Spectrum of protons injected by Dark Matter added to the expected spectrum from standard cosmic ray sources  into a region of shock acceleration at the Galactic Centre compared with data from \textsc{\textsc{KASCADE}}~\cite{0004-637X-612-2-914}, ATIC-2~\cite{Ackermann:2014ula} and CREAM~\cite{2041-8205-714-1-L89}. We assume a DM mass of 100~GeV and an injection rate with efficiency times boost factor $\epsilon \mathcal{B} \sim 10^7$ for supernovae or $\epsilon \mathcal{B} \sim 100$ for injection into the Fermi bubbles. The high-energy cut-off is due to the finite trapping time of the particles in the shock region.}
\label{fig:shock_kas}
\end{figure}

\subsection{Acceleration by the Fermi bubbles} 

 {The so-called Fermi Bubbles are another potential acceleration site in the Galactic centre. Although they may be related to the central supermassive  black hole activity, their detection seem to indicate a large-scale region of shock acceleration of hadrons and/or leptons ~\citep{2014arXiv1407.7905F}. Since, they also coincide with the region where the DM energy density is the highest, cosmic rays injected by DM self-annihilations could get a boost and contribute to the gamma-ray emission from the GC, at high energy. This is particularly interesting if DM is relatively light  ($\sim 1-10$ GeV) \citep{2013PDU.....2..118H} as one might be able to explain both the GeV excess observed in the Fermi-LAT $\gamma$-ray data and the excess  or spectral hardening of cosmic rays reported at higher energy.}

{Assuming that the Fermi Bubbles represent a $\sim 5$~kpc size region of shock acceleration, then the DM annihilation volume is increased accordingly and the injection fraction enhanced, following:}
\begin{equation}
R_{\mathrm{DM}}^{\mathrm{FB}} \sim \epsilon \mathcal{B} \left[ \frac{V_{\mathrm{FB}}}{V_{\mathrm{GC}}} \right] \, \int_{{V_{\mathrm{GC}}}} \mathrm{d}^3 r \, n_{\mathrm{DM}}^2(r) \langle \sigma v \rangle,
\end{equation}
where $V_{\mathrm{FB}}$ represents the volume subtended by the Fermi bubbles and $\epsilon \mathcal{B}$ is a boost factor representing our uncertainty on the acceleration mechanism multiplied by the injection efficiency $\epsilon$. 
\textcolor{black}{The maximum energy to which injected electrons can be accelerated is usually estimated as $E_{\mathrm{max}} \sim e V_s B R_s$, where $R_s$ is the scale-size of the shock.} {\joe{In this case taking $R_s \sim 5$~kpc for the Galactic Centre as an estimate of the Fermi bubble scale, see below, we find $E_{\mathrm{max}} \sim 100$~TeV.}}

\subsection{Expected signatures}
{We show in figure~\ref{fig:shock_kas} a comparison between the proton flux  after acceleration and data from various experiments. Whether DSA occurs in SNRs or near the Fermi bubbles makes no qualitative difference. In both cases, our results indicate the existence of a spectral feature. However this feature is only visible if $\epsilon \mathcal{B} \sim 100$ in the case of the Fermi bubbles and $\epsilon \mathcal{B} \sim 10^7$ in the case of SNRs. Hence we conclude that both are able to accelerate protons injected by DM to high energy but the Fermi bubbles are a more powerful particle accelerator than SNRs when it comes to DM annihilation products. }

{Accelerated annihilation products could also lead to a radio signature if the acceleration mechanism is efficient enough.} We estimate the synchrotron power near either SNRs and the Fermi bubbles using the expression~\cite{2013ApJ...768..106S},
\begin{equation}
\frac{\mathrm{d}\mathcal{W}}{\mathrm{d}\nu} =  \int \mathrm{d} E P(\nu,E) \frac{\mathrm{d}N}{\mathrm{d} E},
\end{equation}
{where $P(\nu,E)$ gives the amount of synchrotron radiation per injected electron and ${\mathrm{d}N}/{\mathrm{d} E}$ is the steady-state spectrum of injected electrons in the shock region. We compare the results to measurements of the `WMAP haze', where an anomalous  radio emission was detected towards the galactic centre around frequencies of about $30 \ \rm{GHz}$. }

\begin{figure}[t]
\centering
\includegraphics[width=0.49\textwidth]{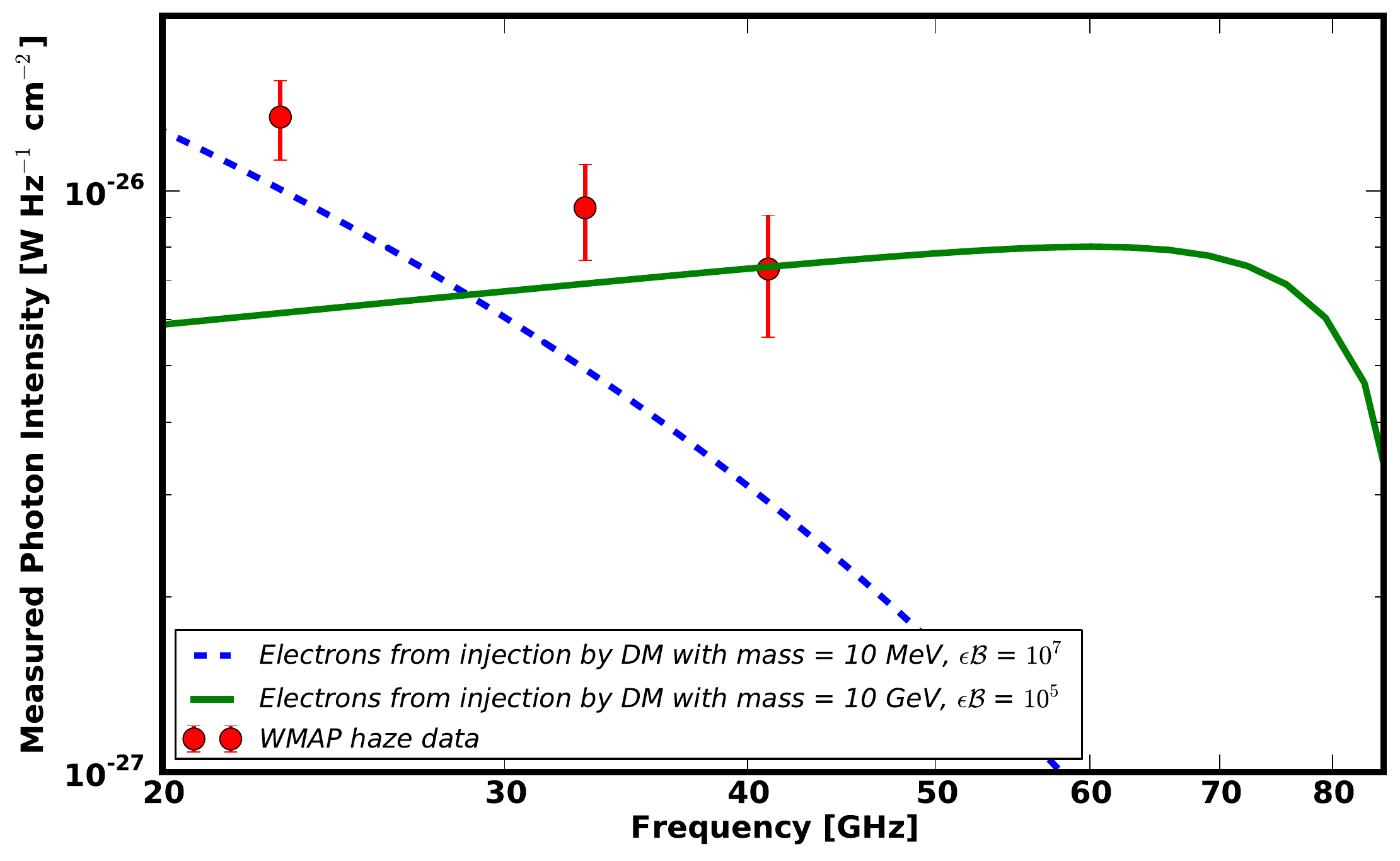}
\caption{Comparison of data from WMAP~\cite{2013A&A...554A.139P,2011PhRvD..83h3517H} {\kumiko (the so-called `WMAP haze': a radio emission detected towards the galactic centre)} to synchrotron emission from electrons injected into regions of shock acceleration corresponding to the Fermi bubbles by dark matter, under two different assumptions for the mass and efficiency times boost factor $\epsilon \mathcal{B}$ of the latter.}
 \label{fig:wmap_haze}
\end{figure}

{The WMAP observations are shown in figure~\ref{fig:wmap_haze}, along with the expected radio signature from DM. While DM can explain the intensity of the WMAP haze, we note that it does not reproduce the haze's spectral shape and cannot therefore be the sole explanation to this anomaly. Besides, whatever the DM mass, the value of $\epsilon \mathcal{B}$ that is needed to make the DM signal visible remains very large.}

{ Furthermore, we note that DM particles heavier than $\sim$ 10 GeV could generate a feature in the radio spectrum that may be visible at higher frequencies and could therefore be of  relevance for the Planck experiment \cite{2013A&A...554A.139P}, which also feature a haze with a similar morphology~\cite{2013A&A...554A.139P,2011PhRvD..83h3517H}. However the dominant dust contribution may prevent signal extraction of a possible DM contribution at high frequencies. Adding polarization to the predicted template might nevertheless help to find further evidence for an "anomalous" synchrotron component. }
\textcolor{black}{For the composition of cosmic rays in the galaxy around and above 1~TeV there will also likely be a sub-dominant composition change of the cosmic rays if there is a bump in the spectrum as in fig.~\ref{fig:shock_kas} from the (highly boosted) accelerated DM.} {\joe {This might for  example manifest itself as a $\bar p $ contribution to the AMS-02 flux.}}

\textcolor{black}{For the specific case of the Fermi bubbles there is an additional interesting point in that the cosmic rays only diffuse through the interstellar medium at a fraction of the Alfven speed. Indeed the diffusion velocity will be less than 100kms$^{-1}$ even in the hot phase of the interstellar medium, and thus it will take more than $10^8$~years for the cosmic rays accelerated in the Fermi Bubble outburst to diffuse to the solar system. Thus the wave of diffusing DM accelerated cosmic rays outlined in this paper will reach the Earth later with a significant delay. Indeed the accelerated  protons and anti-protons propagate outward towards us in the Galaxy disk and halo as a wave. When this wave interacts with e.g. molecular clouds, and if there is an increased energy bump and flux at 1~TeV, then diffuse  gamma ray emission  from the cloud interaction with the CRs will also be generated, although this might not be easily distinguishable from the ambient flux.}

\vspace{0.3cm}
\section{Extra-galactic Signals from Clusters and AGN \label{sec:extra}}

{We  now focus on cosmic rays injected in extragalactic sites where DSA is also meant to occur. We will discuss two acceleration sites: AGN jets~\cite{Rieger:2006md}, where the  dark matter number density is high \citep{Gorchtein:2010xa, 2011PhRvD..84f9903G, 2013PhRvD..88a5024G}, and merging of clusters of galaxies~\cite{Vazza:2013hma}.}

\subsection{AGNs}

{While we do expect electron and proton acceleration in these sites,  we will  focus in this subsection on the acceleration of exotic particles near AGNs.  The acceleration of DM induced cosmic rays that do not have primary astrophysical sources, such as anti-deuterons~\cite{Brauninger:2009pe,Donato:1999gy}, could provide indeed a unique signature of the acceleration of DM annihilation products by shock acceleration.  }

{  It is worth bearing in mind however that measurements of the anti-deuteron flux at extra-galactic energies ($\sim 10^{18}$~eV)  do not currently exist~\cite{Brauninger:2009pe}. Even so we can estimate the fluxes by re-scaling the anti-deuteron flux expected from the galactic centre $\Phi_{\mathrm{GC}}^{\mathrm{AD}}$, see ref.~\cite{Brauninger:2009pe}.  We then need to account for i) the potential increase in DM energy density around  AGNs ($\rho^0_{\mathrm{AGN}}$) compared to the galactic centre $\rho^0_{\mathrm{GC}}$, ii) the different volumes associated with these objects ($V_{\mathrm{AGN}}$ for the AGN versus $V_{\mathrm{GC}}$ for the galactic centre) and iii) the longer distance to the source ($d_{\mathrm{AGN}}$ for the AGN distance versus $d_{\mathrm{GC}}$ for the galactic centre). These different rescaling factors lead to a flux of accelerated cosmic rays near AGN jets of the order}  
\begin{equation}
\Phi_{\mathrm{AGN}}^{\mathrm{AD}} \lesssim \left( \frac{\rho^0_{\mathrm{AGN}}}{\rho^0_{\mathrm{GC}}} \right)^2 \cdot  \left( \frac{d_{\mathrm{GC}}}{d_{\mathrm{AGN}}} \right)^2 \cdot \left( \frac{V_{\mathrm{AGN}}}{V_{\mathrm{GC}}} \right) \cdot \Phi_{\mathrm{GC}}^{\mathrm{AD}}, 
\end{equation}
bearing in mind that such a simple estimate does not account for the efficiency of the AGN shock and other, potentially important, factors.

Taking $d_{\mathrm{GC}} \sim 8$~kpc, $d_{\mathrm{AGN}} \sim 10$~Mpc, $\rho^0_{\mathrm{GC}} \sim$~100~GeVcm$^{-3}$, $\rho^0_{\mathrm{AGN}} \sim 10^{15}$~GeVcm$^{-3}$, $V_{\mathrm{AGN}} \sim 10^{-12}$~kpc$^3$ and $V_{\mathrm{GC}} \sim 1$~kpc$^3$ we find an extra-galactic anti-deuteron flux of $\Phi_{\mathrm{AGN}}^{\mathrm{AD}} \lesssim 10^{-4}$~m$^{-2}$s$^{-1}$sr$^{-1}$GeV$^{-1}$, based on $\Phi_{\mathrm{GC}}^{\mathrm{AD}} \sim 10^{-12}$~m$^{-2}$s$^{-1}$sr$^{-1}$GeV$^{-1}$ for 10~TeV Dark Matter. Even though this upper bound is likely to be overestimated, we have shown that in principle it should be possible to detect extra-galactic anti-deuterons at sufficiently high energies from DM injection near AGNs.

\subsection{Merging clusters}

Merging galaxy clusters also show evidence of particle shock acceleration. A prominent example is the existence of so-called `radio relics', which are extended Mpc sized regions of radio synchrotron emission, present towards the outer edges of merging clusters. The radio emission from these relics is strong. It has a magnitude of order $\sim 10^{24}$~Watts/Hz at $1.4$~GHz~\cite{Vazza:2013hma} and associated  spectra show a clear power-law behaviour, as would be expected from shock acceleration of electrons.
Further evidence is found from the radio spectral index spatial distribution~\citep{2014ApJ...785....1B}, which steepens on either side of the radio relic in some cases. 

{Radio relics are typically observed to have low Mach number shocks of Mpc scale near the cluster virial radius. This poses an injection problem as shocks cannot  easily accelerate  thermal  electrons to sufficiently high Lorentz factors~\citep{2013MNRAS.435.1061P}.  Another problem for standard thermal plasma injection into DSA is the lack of substantial hadronic acceleration signals. 
{{This suggests a ratio in the number of accelerated electrons to protons  that is  of order 0.1, very much higher than the canonical value of 0.01 observed in galactic cosmic rays or the even lower values suggested for SNR acceleration; however there is no evidence for this ratio from the Fermi gamma ray constraints~\citep{2014MNRAS.437.2291V}.  A natural resolution might be  dark annihilation  debris injection which naturally gives  comparable numbers of relativistic protons and electrons, and meets the Fermi constraint.}}}
 \textcolor{black}{The maximum energy to which injected electrons and protons can be accelerated is of order $E_{\mathrm{max}} \sim e B R_s$, where $R_s$ is the scale-size of the shock. In this case taking $R_s \sim 1$~Mpc for the merging cluster shocks we find $E_{\mathrm{max}} \sim 3$~PeV.}

{This strengthens the case for a more exotic injection mechanism, such as DM self-annihilations. Although the number of injected particles is smaller than the ambient population,  cosmic rays injected with an energy of about $\sim 10$~MeV  (corresponding to a DM mass of about 10~MeV, which motivates this particular choice of DM mass) can potentially rehabilitate shocks with a low Mach number.  Since dark matter is expected to exist in abundance in colliding clusters, it may then be possible to observe an enhanced synchrotron radiation emission from DM-injected electrons in these relics. It is worth noticing nonetheless that tight bounds on the DM self-annihilation cross section $\langle \sigma v \rangle$ have already been set, using radio emission from clusters~\cite{Boehm:2002yz, Colafrancesco:2005ji} and one cannot arbitrarily increase the boost factor or the annihilation cross section.}

\begin{figure}[t]
\centering
\includegraphics[width=0.49\textwidth]{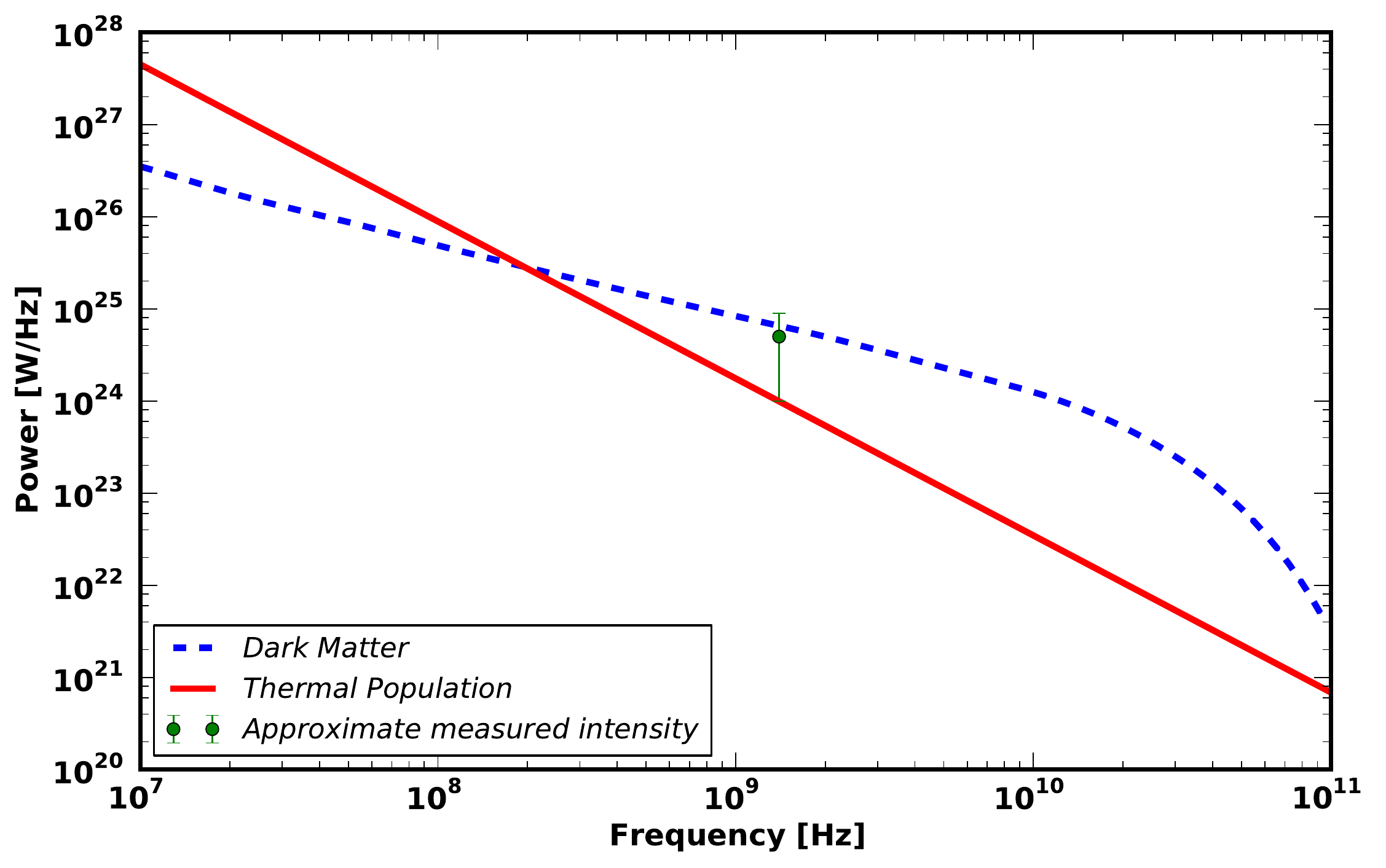}
\caption{Radio spectrum due to synchrotron emission from electrons injected into a region of diffuse shock acceleration by DM particles with a mass of 10~MeV and an annihilation cross section of $\langle \sigma v \rangle \approx 10^{-26}$cm$^3$s$^{-1}$ (and an efficiency times boost factor $\epsilon \mathcal{B} \sim 10^5$), compared with a power-law from the pre-existing `thermal' population. We also show the  spread of measured values of radio emission for the various clusters quoted in ref.~\cite{Vazza:2013hma}.}
\label{fig:radio}
\end{figure}

We base our calculations on the so-called `Toothbrush' relic in the cluster 1RXS J0603.3+4214 ~\cite{2012MNRAS.425L..76B,Vazza:2013hma}. We take a magnetic field of $B = 10 \, \mu$G and  assume that electrons can originate from Dark Matter self-annihilation over a volume of $1$~Mpc$^3$. Assuming that the DM is distributed around the cluster centre according to a Hernquist profile~\cite{2012MNRAS.425L..76B}, and choosing a DM mass of 10~MeV and an efficiency times boost factor $\epsilon \mathcal{B} \sim 10^5$,  we find a synchrotron power of the order $10^{25}$~Watts/Hz at 1.4~GHz, that is of the same order as observations. The resulting spectrum is shown  in figure~\ref{fig:radio} where we display both the DM  and the standard `thermal' electrons contributions. Note that the thermal emission is assumed to obey a power law~\cite{Vazza:2013hma}. 

{While our findings show that a DM signal only becomes visible for very large values of  the $\epsilon \mathcal{B}$ factor, this scenario could be justified by either a greater DM density than expected for the Toothbrush cluster, especially towards its centre, or a larger acceleration region near the relic itself.  However, due to its spectral features, radio observations at higher frequencies (in particular in the $10^9-10^{11}$~Hz range, see  figure~\ref{fig:radio}) could constrain such scenarios. Hence we expect that experiments such as ALMA \cite{alma} could be a useful instrument to test the possible shock acceleration of DM annihilation products. Note that DM particles with a mass of 10~MeV are too light to produce protons via self-annihilation, and so the radio emission from electrons is enhanced without increasing the size of the relativistic proton population.} { This could explain the absence of proton acceleration in such objects, as observed. }

\section{Conclusion \label{sec:conc}}
Diffuse Shock Acceleration at astrophysical sites is the {\joe preferred} {\kumiko process to accelerate} both galactic and extra-galactic cosmic rays. {\kumiko The DSA mechanism typically acts on the ambient thermal population of electrons and nuclei producing particle spectra with a power-law form, as is observed in cosmic ray data. Fermi bubbles, galaxy clusters and Active Galactic Nuclei could be favourable sites for DSA, as they present shock regions, where the signature of particle acceleration has been observed.} {However DSA of thermal cosmic rays cannot always easily account for very high energy cosmic rays.} 

{In this work, we have considered the injection of  Dark Matter annihilation products into  shock regions. Using a simple empirical model we showed that shock acceleration allows DM injection products to be accelerated to energies much higher than the DM mass.  The values of the efficiency times boost factor $\epsilon \mathcal{B}$   is central to all of our estimates and needs to be very high for the DM annihilation products to contribute significantly to the observed cosmic ray spectra.  However this may be realistic due to a boost in the DM number density near acceleration sites.}
\textcolor{black}{Even in the strongest collisional shock one might expect to have in a cluster merger such as  the Toothbrush,  the Mach number is moderate and any dark matter density enhancement can only be a factor of around  four. However a much larger density enhancement could occur towards the galactic centre  or even more plausibly towards supermassive black holes such as those in M87 and Cen A, due to the presence of a central massive black hole~\cite{Gondolo:1999ef}. The result may be  an enhancement of  the dark matter density by many orders of magnitude due to the presence of a spiked density profile. This may therefore account for the large boost factors used in this work. }{\joe{The jet in  M87 offers an especially attractive environment for shock-boosting of DM annihilation debris \cite{2015arXiv150500785L}. Indeed  even without such boosting, a gamma ray signal is plausibly detectable from jet-DM scattering in the case  of Cen A \cite{2013PhRvD..88a5024G}.}}{\joe{   One can even imagine ultraheavy, non-thermally produced, DM particle annihilation products that might be jet-boosted to ultrahigh cosmic ray energies in the vicinity of dark matter-enhanced density spikes around active SMBH  by jet acceleration mechanisms as in \cite{2015arXiv150506739C}.}} However in all cases our proposed signal would also have to compete with the prompt signal from DM-annihilation products outside the shock region.

{ More conservatively, however, we find that using simplifying assumptions, protons injected near SNRs in the Galactic Centre would lead to an observable signature if  $\epsilon \mathcal{B} \sim 10^7$ while this factor would be about $100$ if the origin of the acceleration is the kpc-sized Fermi Bubbles, thus showing that Fermi bubbles are an efficient accelerator of DM annihilation products. 
Similar values are obtained for the acceleration of electrons but it is worth noticing that electrons could in addition induce a potential contribution to the WMAP haze.}
 
{Injection of cosmic rays by DM annihilations near extra-galactic sites where DSA is happening could also lead to interesting signatures. The observation of  exotic particles at extra-galactic energies ($\gtrsim 10^{18}$~eV) without any astrophysical primary counterpart, such as anti-deuterons, would be an important signature of DM injection at shocks. In addition  electrons injected by  DM annihilation could give rise to an important contribution in radio relics present around merging clusters, for cross-sections around $\langle \sigma v \rangle \sim 10^{-26}$~cm$^3$s$^{-1}$ and 10~MeV particle masses, as can be seen in figure~\ref{fig:radio}. }{
Our modelling of the shock acceleration and the DM-injection of electrons and protons is admittedly very simplistic. More sophisticated simulations, which include the propagation of the  DM-annihilation products to the shock regions as well as more accurate energy losses and magnetic diffusion processes are required to confirm our values of the flux. However we expect the general spectral morphology (namely a power-law with a low energy cut-off) to remain an interesting spectral feature for future cosmic ray searches.

\section*{Acknowledgement}
This research has been supported in part by the Balzan Foundation via  Johns Hopkins University. JHD is grateful to the to the Physics and Astronomy Department at the Johns Hopkins University for their hospitality. The research of JS and JHD has also been supported at IAP by  ERC project 267117 (DARK) hosted by Universit\'e Pierre et Marie Curie - Paris 6 and for JS at JHU by National Science Foundation grant OIA-1124403. KK is supported by Sorbonne Universit\'es.

\appendix
\section{Details of the Empirical Model \label{app1}}
In this appendix we list the main steps used to calculate the spectra of electrons or protons after injection by Dark Matter into regions of shock acceleration.
\begin{enumerate}
\item \emph{Initial spectrum}: The cosmic rays produced by Dark Matter enter the shock and accumulate according to 
\begin{equation}\label{eq:DMinj}
\frac{\mathrm{d}n}{\mathrm{d}E}(t + \Delta t, E) = \frac{\mathrm{d}n}{\mathrm{d}E}(t,E) + \epsilon  n_{\mathrm{DM}}^2 \langle \sigma v \rangle f_{\mathrm{DM}} (E,m) \Delta t
\end{equation}
where $n_{\mathrm{DM}}$ is the Dark Matter number density, $\langle \sigma v \rangle$ is the annihilation cross section and $f_{\mathrm{DM}} (E,m)$ is the normalised spectrum of injected electrons or protons which depends on the DM mass $m$. Throughout this work we assume that $\langle \sigma v \rangle \sim 10^{-26}$~cm$^3$s$^{-1}$ i.e. the commonly chosen `thermal' value.
We make the assumption that DM far from the shock can still inject particles which reach the shock region, given a suitable diffusion model. Hence throughout this work we integrate over the entire DM distribution when calculating the injection rate, but correct for the smaller volume over which shock acceleration takes place.

In this work we assume that $f_{\mathrm{DM}} (E,m) \propto \delta(E - m)$ for electrons i.e. that all injection occurs at the DM mass.
Note that we assume the electron cooling time is much shorter than their diffusion time
(in the case of pairs)
but this is not the case for protons for which the cooling time by photopion or $e^+ e^-$ pair production is probably much longer than the diffusion time
Indeed as can be seen in figure~\ref{fig:shock_nonrel} the resulting spectrum depends strongly on the value of $m$.
\item \emph{Energy losses}: By the end of the time-step the particles trapped in the shock, including those injected at the start of the time-step, will have lost energy. 
Since we are interested in GeV energies and above, we assume that the electrons lose energy through Inverse Compton scattering with diffuse light or synchrotron radiation~\cite{2013ApJ...768..106S}, while protons additionally lose energy to $e^+ e^-$ pair production~\cite{Kelner:2008ke,Achterberg:1999vr}. Hence for each particle with index $i$ the energy losses are modelled as,
\begin{equation}
E_i(t + \Delta t) = E_i(t) - b_{\mathrm{loss}} E_i^2(t) \Delta t
\end{equation} 
where $b_{\mathrm{loss}}$ is the total loss rate in units of GeV$^{-1}$s$^{-1}$.
\item \emph{Shock crossing or escape}: The particles only remain trapped in the shock region for a finite amount of time. At the end of each time-step a certain fraction of particles will leave the down-stream region, according to the time-scale for particle trapping $T_{\mathrm{trap}}$.
Hence there are $\left(1 - \mathrm{exp} \left[ - {\Delta t}/{T_{\mathrm{trap}}} \right] \right)$ particles which leave the down-stream region.

For these particles there are two possible outcomes: they can either cross the shock again with a probability $\beta$ and return to being trapped, but in the up-stream region, or they can escape the shock region, with probability $1 - \beta$. For example if a particle scatters off an irregularity in the magnetic field, the angle of deflection determines whether it crosses the shock or escapes the shock region entirely~\cite{Achterberg:2001rx}.

For the particles $j$ which cross the shock their energies are incremented according to,
\begin{equation}
E_j(t + \Delta t) = E_j(t) + \delta \cdot E_j(t)
\end{equation} 
where $\delta$ is the fractional energy gain per crossing. Indeed the values of $\beta$ and $\delta$ set the spectral index according to $\alpha = (\mathrm{ln} \, \beta / \mathrm{ln} \, \delta) - 1$. The fractional energy gain for a non-relativistic shock is $\mathrm{ln} \,  \delta = U / c$ where $U$ is the shock velocity and $c$ is the speed of light. For a shock with a Mach number $\mathcal{M} = 10$ we have $\mathrm{ln} \, \delta \sim 10^{-5}$ and so for example to get $\alpha = 2$ (close to the expected injection spectral index at astrophysical sources~\cite{Rieger:2006md,Achterberg:2001rx}) we need $\mathrm{ln} \, \beta \sim - 10^{-5}$ i.e. a small energy gain per shock crossing, but a high probability that a particle will cross the shock multiple times. The spectrum will also steepen when energy losses are taken into account.
\end{enumerate}

\end{document}